\documentclass{aa}
\usepackage{graphicx} 

\usepackage{natbib} 
\bibpunct{(}{)}{;}{a}{}{,}

\begin{document}

 \title{A weak upper limit for axion-related continuum polarization signal at 630 nm}

   \author{C. Beck\inst{1} }
        
   \titlerunning{A weak upper limit for axion-related continuum polarization}
  \authorrunning{C. Beck}  

   \institute{National Solar Observatory, Boulder/CO}
 
\date{Received xxx; accepted xxx}
\abstract{Axions could be a component of the dark matter and are of great interest for particle physics. If present, their interaction with magnetic fields in the solar photosphere should produce a 
  polarization signal of real photons, whose amplitude depends quadratically on the magnetic field strength.}{I search for possible traces of axion-induced
  low-energy polarization signal in the visible wavelength range.}{I investigate the polarization signal at a continuum wavelength in the visible near 630 nm. I use an observation of a sunspot and the nearby solar active region to study the dependence of the
  continuum polarization signal on magnetic field strength. I
  correct the continuum polarization signal for the polarization introduced by
  the telescope due to oblique reflections. I reproduce the corrected
  polarization spectra using a spectral synthesis of solar model atmospheres
  that include the Zeeman effect in the presence of magnetic fields, but no
  axion-related processes. These synthetic spectra serve as a reference for
  the amount of polarization from a known source.}{I
  find no trace of an enhanced polarization signal in continuum polarization
  above a polarization level of around 0.05 \%. The continuum polarization
  and its spatial variation across the field of view at this level can be
  traced back to either the Zeeman effect or residual cross-talk between
  intensity and polarization.}{I provide a weak upper limit for the existence
  and interaction of axions with solar magnetic fields by proposing that any
  model of solar axion generation and interaction should not lead to
  polarization effects above  0.05 \% of the continuum intensity at a wavelength of 630\,nm, i.e., not more than 5 out of 10000 photons should show polarization 
  due to axion-related processes. Using estimates of photon and axion flux rates, I derive an upper limit of $g_{a\gamma} < 700 \cdot 10^{-10}$ GeV$^{-1}$ for the axion-photon coupling constant $g_{a\gamma}$.}
\maketitle

\section{Introduction}
{\bf Comment: The last time this publication was edited was Aug 21, 2008, according to the time-stamp. It was never submitted because of the primarily negative result. At that time, there was a ``solar axion telescope'' at CERN. I did not look up now if that led to any results.}\\$ $\\

Axions are hypothetical particles that could constitute a component of the dark matter \citep[see, e.g.,][and references therein]{andriamonje+etal2007}. Axions can transform into real photons in the presence of magnetic fields with a probability $p_{a\rightarrow\gamma}$ given by 
\begin{equation}
p_{a\rightarrow\gamma} = \left(\frac{g_{a\gamma}B}{q}\right)^2 \cdot \sin^2
\left(\frac{qL}{2}\right) \;, \label{eq1}
\end{equation}
where $q=m_a^2/2E$ is the photon-axion momentum difference, $g_{a\gamma}$ the axion-photon coupling and $L$ the extension of the magnetic field with field strength $B$ \citep[taken from][]{andriamonje+etal2007}. An expansion of the sinus term for $qL << 1$ leads to $p_{a\rightarrow\gamma} =1/2 (g_{a\gamma}BL)^2$ \citep[][ Eq.~(12)]{cameron+etal1993}. 

The Lagrangian for the interaction of axions and photons contains a scalar
product with the magnetic field vector {\bf B}, and hence, an anisotropy is
introduced in the directions parallel or perpendicular to the magnetic field
direction. This leads to a rotation of linear polarization or the introduction
of in general elliptically polarized light \citep[both linear and circular polarized components;][]{cameron+etal1993}. Axions are assumed to be generated in
the solar fusion core. Because of their weak interaction with matter, they can
reach the solar surface with little hindrance. The solar surface is, however,
known to contain magnetic fields of considerable strength (up to 0.3 T in the
center of sunspots) with a large spatial extension (several 10000 kms). A
possible interaction of axions generated in the solar core with the surface magnetic fields then should introduce a polarization signal in excess of other known
sources of polarization, like the Zeeman effect in solar absorption lines. The
additional polarized photons should not be related to specific wavelengths and
thus could best be seen in continuum wavelength ranges, where the black-body
solar radiation is unpolarized due to the incoherent emission of thermal photons. Any excess continuum polarization in the presence
of strong magnetic fields should then also have a quadratic dependence on the
field strength (cf.~Eq.~\ref{eq1}). 

To investigate the topic, two quantities have to be determined: the
magnetic field strength in the solar atmosphere and the continuum
polarization. To this extent, I analyzed a polarimetric observation in the
visible wavelength range at 630 nm of an active region on the surface of the Sun. The polarization signature of the magnetic fields due to the
Zeeman effect was used to determine the field strength in a two-dimensional (2-D) field of view covering a small and a big sunspot. The continuum polarization was derived
from the same data set, which was re-calibrated without the correction for the
telescope and the continuum polarization that is part of the usual data
reduction. The continuum polarization signal was investigated first in
relation to the one predicted by the geometrical model of the telescope that
generates continuum polarization by oblique reflections. As second step, the
continuum polarization due to the Zeeman effect was studied. Any polarization
in excess of these two known sources could be indicative of axion-related processes.

I shortly introduce the telescope, the polarimeter and the data set
(Sect.~\ref{teletal}). The data reduction procedure is discussed in some
detail in Sect.~\ref{stand}, because it is closely related to the question of
cross-talk, i.e., the unwanted transformation of polarization states into each
other. In Sect.~\ref{ana}, I explain the data analysis
procedure and the found behavior and properties of the continuum
polarization. The findings are summarized and discussed in Sect.~\ref{summ}. 
\begin{figure}
\centerline{\resizebox{3.cm}{!}{\includegraphics{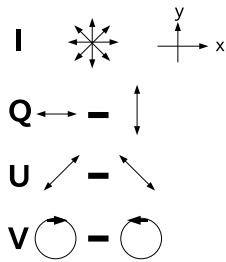}}}
\caption{Graphical sketch of the four Stokes parameters $IQUV$.\label{sstokes}}
\end{figure}
\section{Telescope, polarimeter \& observation\label{teletal}}
In the following study, I will describe the polarization state of the light in
the \emph{Stokes formalism} \citep[e.g.,][]{shurcliff1966,collett1992}. The polarization
state is quantified uniquely by four parameters, commonly combined to the
Stokes vector {\bf S} = ($I, Q, U, V$). The four entries correspond to the
total intensity, $I$, the difference of the amount of linear polarization along
the $x$ and $y$ axes, $Q$, the difference of linear polarization at $\pm 45$
degree, $U$, and the difference of right and left circularly polarized light,
$V$. Figure \ref{sstokes} shows the commonly used graphical representation of
the Stokes parameters. 
\subsection{The German Vacuum Tower Telescope (VTT)\label{tell}}
\begin{figure}
\centerline{\resizebox{3.cm}{!}{\includegraphics{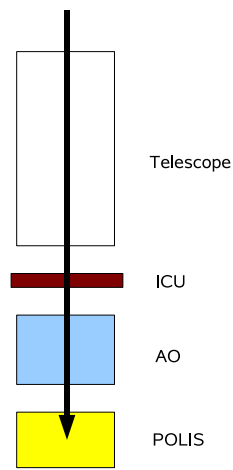}}}
\caption{Schematical light path at the VTT. The light passes the telescope and
  the AO system before it reaches the polarimeter POLIS. The
  ICU can be inserted right behind the
  telescope to calibrate the remaining light path.\label{schema}}
\end{figure}

The VTT is a coelostat telescope at the Observatorio
del Teide in Iza{\~n}a, Tenerife (Spain). Two flat 70-cm mirrors are located on top of an eleven-floor
tower. The first coelostat mirror, C1, catches the sunlight, the second
coelostat mirror, C2, deflects it vertically down into 
an evacuated steel tube. Inside the steel tube, the main imaging mirror, M3,
with a 46-m focal length reflects the light $\sim$20 m upwards again to a
folding mirror, M4. The folding mirror reflects the light down again. The
light beam passes the exit window of the evacuated part after again  $\sim$20 m. The telescope focus finally is located in the 1st floor of the building. The Instrumental Calibration Unit (ICU) for a calibration of polarimeters is located right behind the exit window. In 2003, the Kiepenheuer-Institute adaptive optics (AO) system was installed to improve the spatial resolution by a real-time correction of the wavefront distortions introduced in the earth atmosphere. The corresponding optics are
located between the ICU and the polarimeter, as sketched in Fig.~\ref{schema}.
\subsection{The Instrumental Calibration Unit (ICU)}
The ICU consists of a linear polarizer followed by a retarder. With the ICU,
known polarization states can be generated. The instrument response to these
known polarization states is measured, and the linear response function can be
determined from a comparison between input (known state) and output (measured
polarization). The method is described in detail in
\citet{beck+etal2005b} and \citet{beck2006}. The application of the inverse of the response function to the observed Stokes vector corrects all changes of the Stokes vector that were introduced by optics behind the ICU. 
\subsection{The Polarimetric Littrow Spectrograph (POLIS)}
One of the polarimeters at the VTT is the Polarimetric Littrow
Spectrograph. The instrument is designed for measurements of the Stokes vector
at two {\em fixed} wavelength ranges in the visible and near-UV (630 nm, 396.8
nm). The instrument uses a rotating waveplate for polarization modulation. The
modulated polarization is transformed into a modulated intensity level by
polarizing beamsplitters. The polarization state can be reconstructed from the
measured variation of the intensity level. The instrument is described in
detail in \citet{beck+etal2005b}. I used only the data from the channel at
630 nm, since the near-UV channel was not functional during the observations.
\subsection{Data set}
\begin{figure}
\centerline{\resizebox{8cm}{!}{\includegraphics{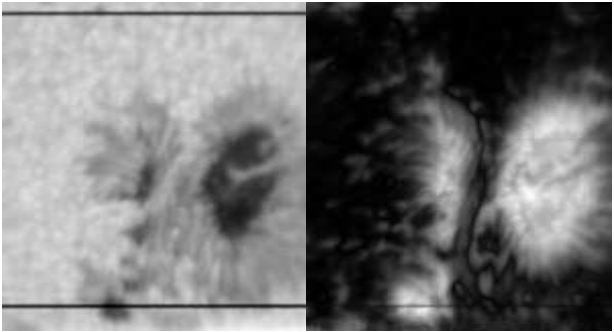}}}
\caption{Overview over the observed field of view. Left: continuum
  intensity near 630.3 nm. Right: circular polarization signal.\label{fig1}}
\end{figure}

\begin{figure*}
\centerline{\resizebox{14.6cm}{!}{\includegraphics{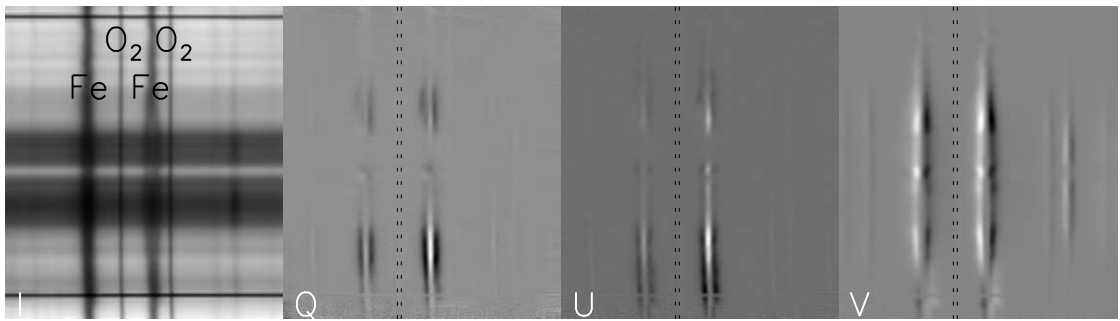}}}
\caption{Spectrum along the slit. {\em Left to right}: $IQUV$. Wavelength increases
  from left to right in each Stokes parameter, the position along the slit
  from top to bottom. The dashed vertical lines outline the continuum window
  used throughout the following.\label{slit_spec}}
\end{figure*}

For the investigation of the possible existence of axion-induced continuum
polarization, an observation taken on the July 6, 2005 was
used. Figure \ref{fig1} shows an overview on the observed field of view
(FOV) in intensity  and circular polarization. The FOV covers a
big sunspot in the right half, with a clearly discernible dark umbral core and
a fully formed penumbra all around the umbral core. A smaller sunspot was
located left of it that had a penumbra only in the left-upper part. The {\em two horizontal black} lines in the intensity map of Fig.~\ref{fig1} are thin wires that were put across the slit to mark the  FOV of the CCD camera. Black
areas in the circular polarization map can be assumed to be free of magnetic
fields to first order. To obtain the 2-D map of the solar surface,
the 40 micron wide entrance slit of the spectrograph was stepped across the
solar image in the focal plane (``scan''). For each step of the scan, the
Stokes vector $IQUV$ of the solar spectrum from around 630.05 nm to 630.4 nm
was obtained (cf.~Fig.~\ref{slit_spec}). The scanning proceeded from left to right, i.e.,the time increases also from left to right. The data set was taken during the time
reserved for the International Time Program (ITP), where several telescopes on
the Canary islands of Tenerife and La Palma were used in coordinated
observations. 


\section{Standard data calibration procedure\label{stand}}
The VTT has not been designed for polarimetry in its very beginning. The
telescope is not axi-symmetric and has several oblique reflections on
mirrors. Already the first mirror, C1, changes the polarization state
significantly and also introduces a time-dependent continuum polarization
level at all wavelengths. All oblique reflections anywhere downstream in the light path change again the polarization state. The ICU
allows to calibrate the instrument response for all changes of the Stokes
vector behind it, but not in front of it. The standard data calibration procedure thus consists of two main steps: a correction for all changes behind (Sect.~\ref{polcal}) and in front of the ICU (Sect.~\ref{telcal}).
\subsection{Instrument calibration\label{polcal}}
With the instrument calibration, the polarization measurements are corrected
for all changes of the polarization state behind the
ICU. The polarimeter response function is determined from the comparison of
the polarization states generated with the ICU and the corresponding Stokes
vector observed with the instrument. For the instrument calibration, it is not
necessary to model the contribution of, e.g., each mirror separately. Only the
final net effect on the polarization is determined. The instrument calibration
is very accurate, because the ICU properties are well known and a linear
relation between ICU state and measured state can be taken for granted. The
method is  described in \citet{beck+etal2005b}. The instrument response is
determined once per day and shows little to no time dependence, because it
reflects the geometry of the light path behind the ICU that is not changing.

\subsection{Telescope calibration\label{telcal}}
No direct calibration is possible for the telescope. To remove the
polarization changes due to the telescope, a geometrical model of the
telescope is used. It contains the actual light path inside the telescope (C1
to exit window) at any given moment of time. The properties
(i.e., the complex refractive index) of the aluminum-coated mirrors were
determined by different methods \citep{beck+etal2005a,beck2006}. With the refractive index of the mirrors, the
changes of the polarization state due to the oblique reflections on mainly C1 and
C2 can be calculated and removed. The telescope calibration is less accurate
than the instrument calibration because aging and other degradation (dust, dirt) of the mirror coatings is not taken into account and because the determination of the complex refractive index is only possible within limits. The
telescope model has to be calculated for each moment of time with the location
of the Sun on the sky and the position and orientation of the mirrors C1 and
C2. 
\subsection{Removal of residual cross-talk}
For studies of the polarization signal due to the Zeeman effect, the continuum
polarization is commonly assumed to be zero. As both the instrument
calibration and the telescope calibration have limitations, the continuum
polarization level in calibrated data is usually non-zero. This spurious polarization is produced by cross-talk, i.e., the transformation of one Stokes parameter into another. The main effect is a transformation of intensity, $I$, into the
polarization signals $QUV$, because for solar light the intensity $I$
is commonly one order of magnitude larger than the polarized fraction,
$\sqrt{Q^2+U^2+V^2}$. In the standard data reduction, any residual continuum
polarization is thus {\em forced} to zero by subtracting a corresponding
fraction of the intensity signal. 

\section{Data analysis\label{ana}}
For the study of the possible continuum polarization induced by axion-related
processes, the data set was reduced two times: once the usual standard
calibration including the telescope calibration and the removal of residual
cross-talk was applied, in the second case only the instrument calibration was
applied. The second data sets thus contains both the continuum polarization
induced by the telescope, and every other effect that could generate
polarization in continuum wavelength ranges. From the observations, one
obtains the four Stokes parameters of the spectrum in a small wavelength range
(Fig.~\ref{slit_spec}) for each location of the slit during the scan. The
spectral range contains two strong solar spectral lines of iron (Fe), and two strong absorption lines of oxygen from the Earth's
atmosphere (O$_2$). The teluric lines do not show any polarization signal or
any measurable Doppler shifts due to velocities. The solar lines show the
effect of solar magnetic fields due to the Zeeman effect. The spectral lines
split into three components in the presence of strong fields (middle of the
FOV). Depending on the orientation of the magnetic field lines to the line of
sight (LOS) between observer and target, the polarization signal shows up stronger
in $Q,U$, or $V$ ({\em longitudinal} or {\em transversal} Zeeman effect). 
\begin{figure}
\centerline{\resizebox{8cm}{!}{\includegraphics{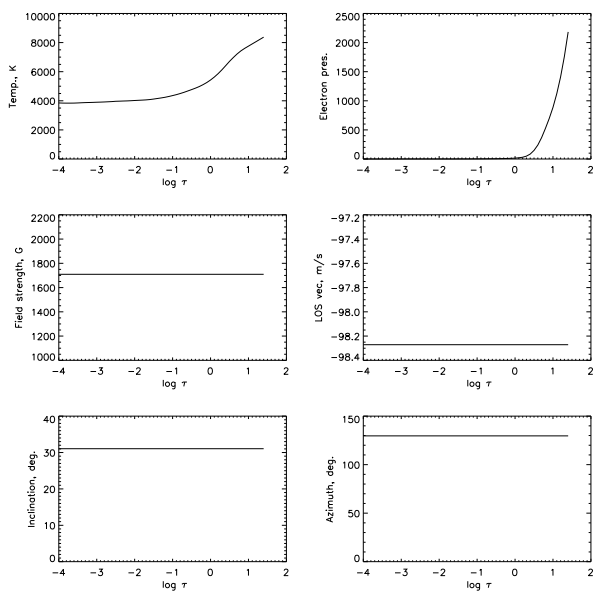}}}
\caption{Solar model atmosphere for the generation of synthetic spectra for
  one spatial location. {\em Left column, top to bottom}: temperature, field
  strength, field inclination. {\em Right column}: electron pressure (=density),
  velocity, field azimuth. \label{figmodel}}
\end{figure}
\begin{figure*}
\centerline{\resizebox{8cm}{!}{\includegraphics{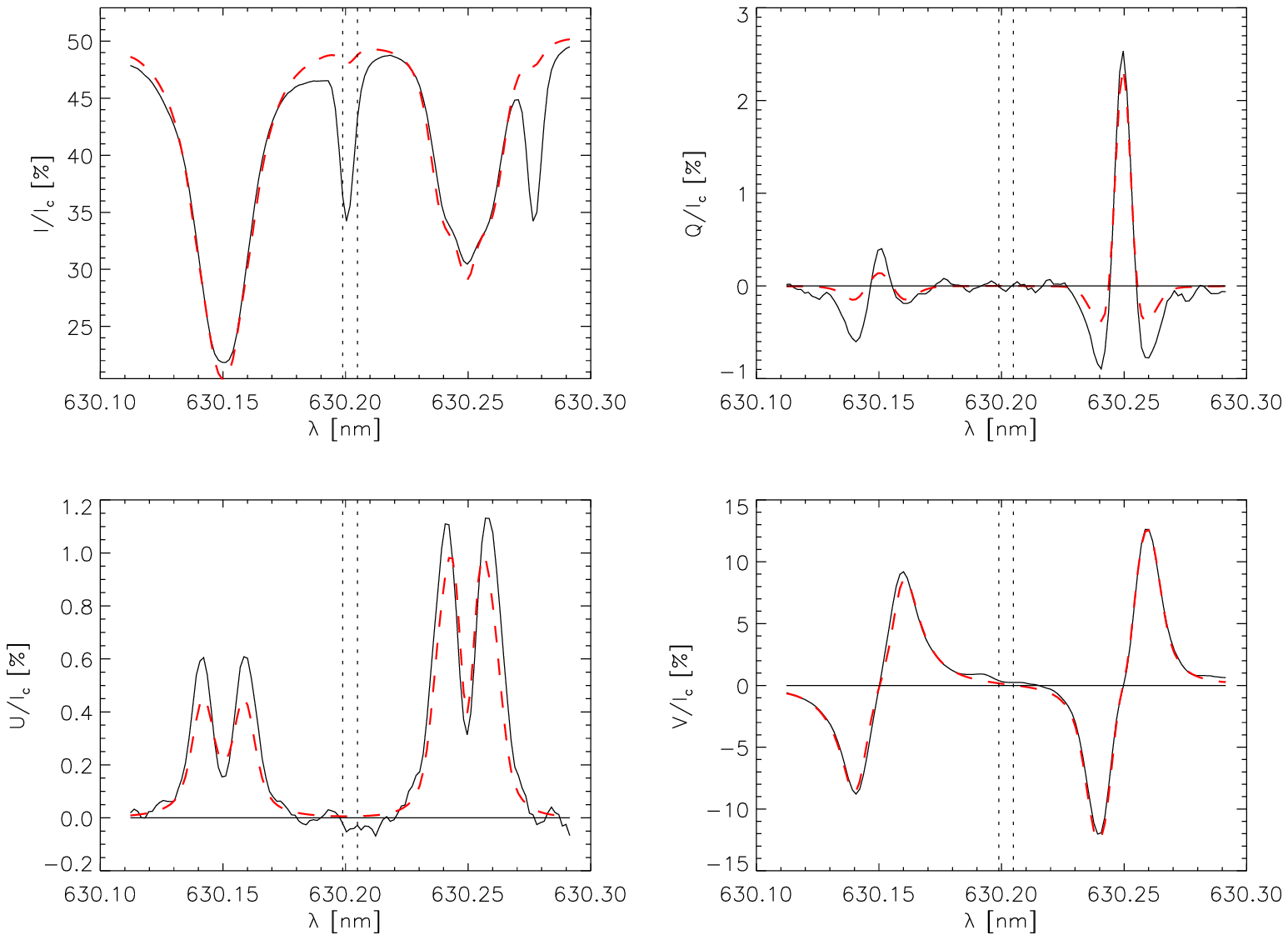}}\resizebox{8cm}{!}{\includegraphics{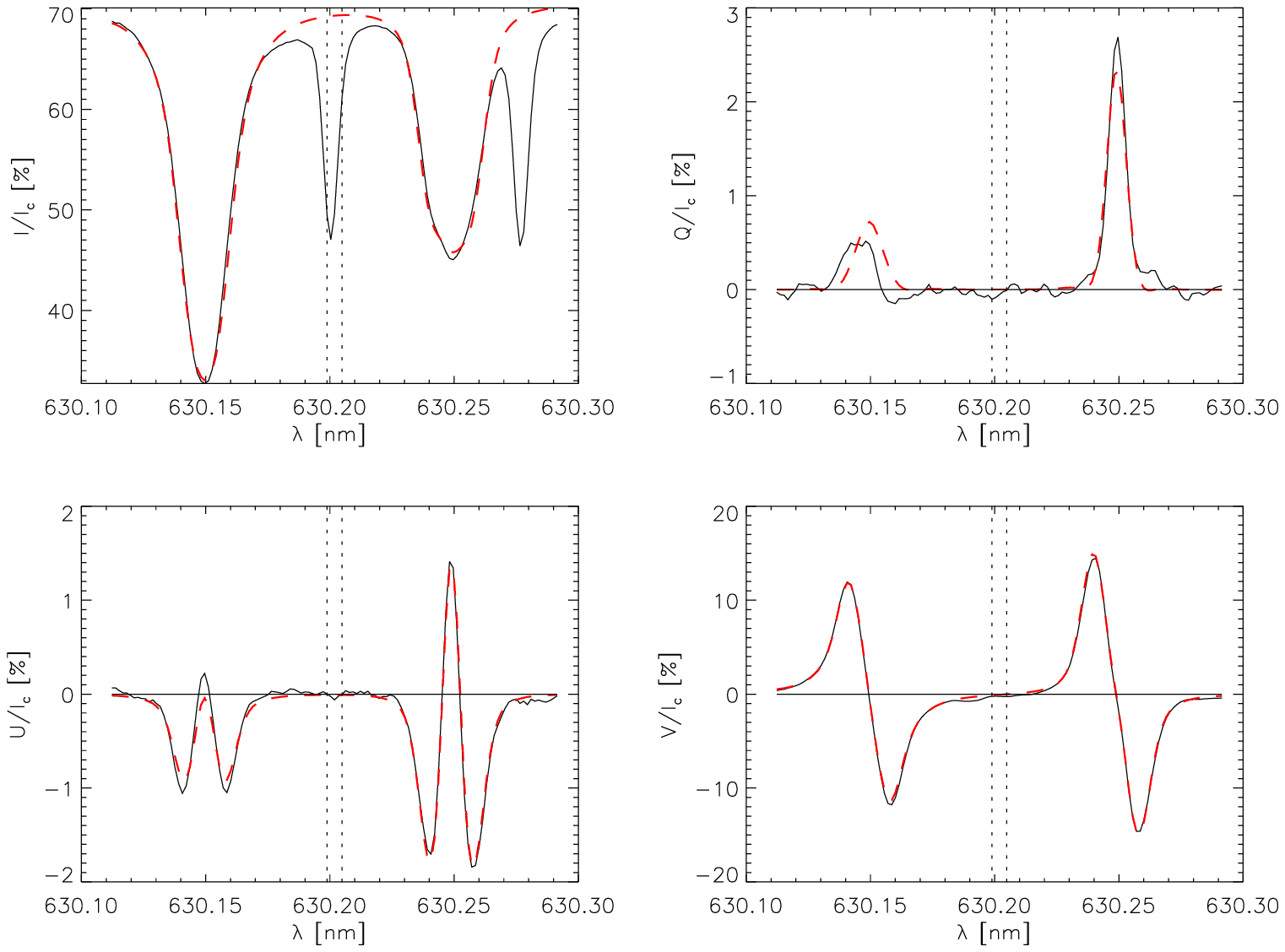}}}
\caption{Observed and synthetic spectra observed on two different spatial
  locations. {\em Clockwise, starting left top}: Stokes $IQVU$. {\em Black solid lines} show the observed profiles, {\em red dashed} the synthetic spectra. The {\em dotted vertical lines} outline the continuum window. Note the opposite polarity in the two Stokes V plots ({\em bottom row, 2nd and 4th column}): the order of maxima and minima
  of the profile is exchanged, indicating magnetic fields (anti-)parallel to the line of sight. \label{figspec}}
\end{figure*}
\subsection{Synthetic spectra}
To obtain the field strength in the observed FOV, synthetic spectra
were generated. The procedure is as follows: one assumes a model for the
solar atmosphere. The solar model atmosphere (cf.~Fig.~\ref{figmodel})
contains the thermodynamic properties (temperature, density, velocity),
and the magnetic field properties (field strength, field inclination (= angle
between field and LOS), field azimuth (= direction of field in the
plane perpendicular to the LOS). The model atmosphere is given on
grid points in optical depth, log $\tau$, instead of a scale in km, because
this is fully sufficient for a synthesis of spectra. The spectra resulting
from the model atmosphere are calculated. However, they will first not match
well to the observed spectra. The properties of the model atmosphere are then
iteratively modified in a least-square-fit, until the synthetic spectra match
the observed ones.  A number of such ``inversion\footnote{From spectra to
atmosphere is the inverse problem to the generation of the spectra in nature.}'' codes are available. I used
the SIR code \citep{cobo+toroiniesta1992}. The inversion is executed
independently for each spectrum (=row of pixels on the CCD) along the slit in every scan step.

Figure \ref{figspec} shows
a comparison of observed and synthetic spectra for two locations in the field
of view. The teluric O$_2$ lines were not included in the synthetic
spectra. From the model atmosphere that yields the best-fit spectra one
obtains the field strength and field orientation for the solar location where
the observed spectra came from. Note that these synthetic spectra {\em cannot
  contain} any continuum polarization possibly induced by axions, only
the polarization due to the Zeeman effect. The continuum polarization induced
by the telescope was also removed prior to the inversion. The synthetic spectra
serve as a reference for the continuum polarization signal of known
origin, i.e., from the Zeeman effect.
\subsection{Continuum polarization induced by telescope and Zeeman
  effect\label{teepol}}
\begin{figure}
\centerline{\resizebox{8cm}{!}{\includegraphics{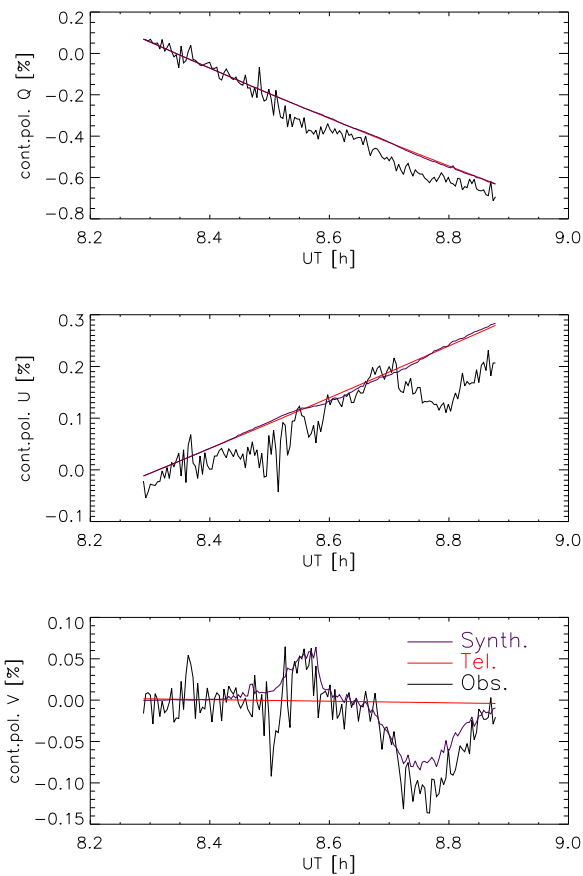}}}
\caption{Continuum polarization induced by the telescope and the
  Zeeman effect, averaged along the slit. {\em Top to bottom}: Stokes $QUV$. {\em Black}:
  observed continuum polarization. {\em Red}: continuum polarization predicted from
  the telescope model. {\em Purple}: telescope polarization plus the continuum polarization in the synthetic spectra.\label{indpol}}
\end{figure}

Before turning to the spatial variation of the continuum polarization signal,
I investigated the polarization level inside the continuum window marked in
Fig.~\ref{figspec}, averaged along the slit in each scan step. I used the
second reduction of the data set where the telescope and residual
cross-talk correction was not applied. Figure \ref{indpol} shows this average
continuum polarization as a function of time during the scanning. The {\em black line} shows the observed continuum polarization. The {\em red line} is the
polarization that is predicted by the telescope model. The {\em purple line} shows the addition of the polarization in the synthetic spectra plus that of the telescope. The observed and predicted continuum polarization agree satisfyingly,
the slope with time predicted by the telescope model is well
reproduced. Examining the curve for Stokes $V$, the locations of the small and
the big sunspot can be recognized: the small spot was scanned at around 8.55
UT, the big sunspot at around 8.75 UT, producing an increased and decreased continuum polarization, respectively. The deviation in opposite
 directions is systematic and has an easy explanation. 

\begin{figure}
\centerline{\resizebox{8cm}{!}{\includegraphics{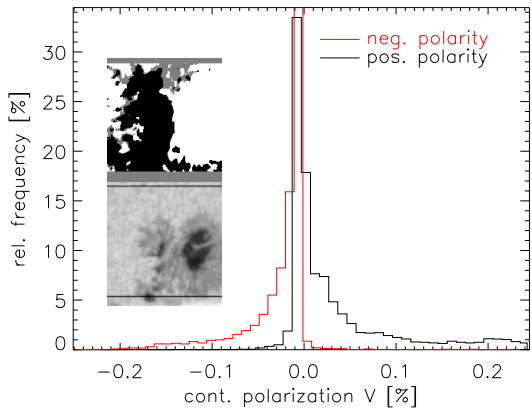}}}
\caption{Histograms of the continuum polarization for locations with positive
  ({\em black}) and negative polarity ({\em red}). Inserted are 2-D maps of polarity  ({\em upper}) and intensity ({\em lower}) in the left part. {\em Black} and {\em white} denote the opposite polarities, {\em grey} indicates no polarization signal. \label{polarity}}
\end{figure}

The circular polarization signal changes its so-called \emph{polarity}, when
the magnetic field is parallel or anti-parallel to the LOS. In the
spectra, this change is visible as the exchange of the order of maximum and
minimum Stokes $V$ polarization signal (cf.~Fig.~\ref{figspec}). As the Zeeman
induced polarization signal extends into the continuum window used to derive
the continuum polarization, the change of the polarity leads to a change of
the sign of the continuum polarization. This is demonstrated also in
Fig.~\ref{polarity}. The figure shows histograms of the continuum polarization
of all locations with positive and negative polarity, respectively. Positive
(negative) polarity is directly related to a positive (negative) continuum
polarization signal. In total, the continuum polarization of the synthetic
Stokes $V$ spectra ({\em purple curve} in Fig.~\ref{indpol}) shows the same behavior
as the observed polarization which implies that this variation is induced by the Zeeman effect. This is not the case for the
linear polarizations $Q$ and $U$: the polarization in the synthetic spectra
does not match the observed polarization. The observed deviation from
the prediction of the telescope model is much larger than that in the
synthetic spectra. This observed variation of the linear polarization is thus
not due to the Zeeman effect.
\subsection{2-D maps of the continuum polarization\label{2dmaps}}
\begin{figure*}
\begin{minipage}{14cm}
\centerline{\includegraphics{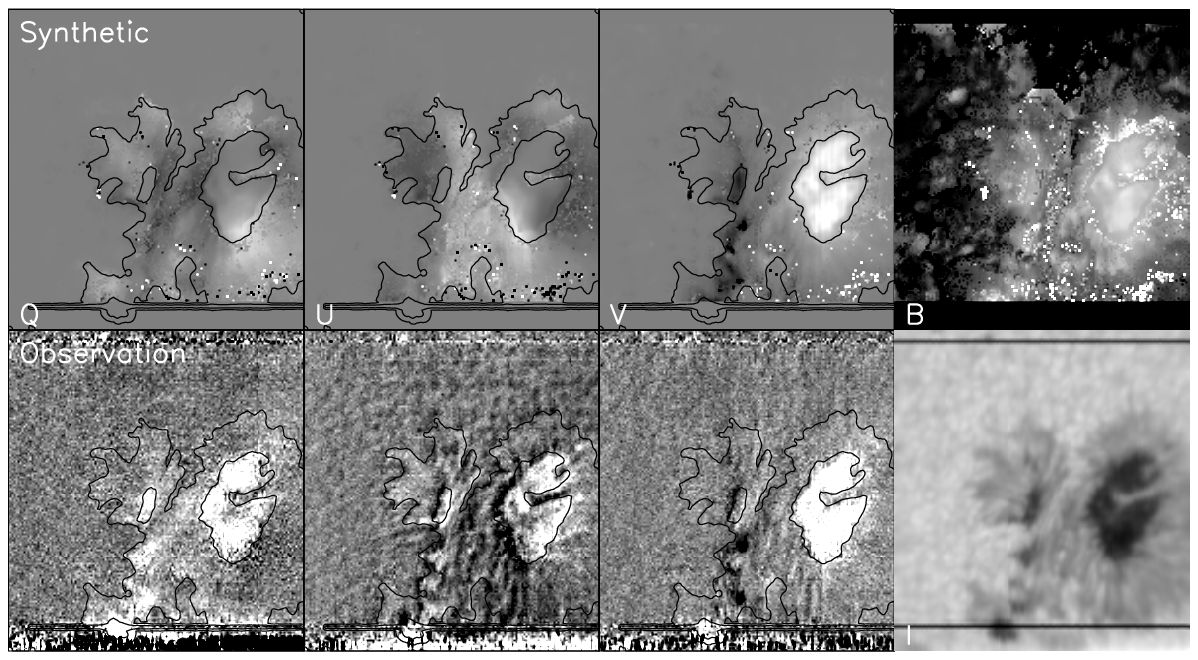}}
\end{minipage}\begin{minipage}{4cm}
\includegraphics{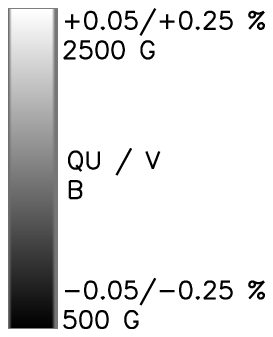}\\
\includegraphics{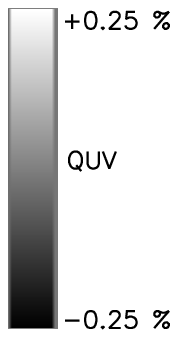}
\end{minipage}
\caption{2-D map of continuum polarization in the synthetic ({\em top row})
  and observed spectra ({\em bottom row}). {\em Left to right}: $QUV$. {\em Last column, top}: field strength; {\em last column, bottom}: Stokes $I$. The {\em black contours} outline the umbra and penumbra of the big and small sunspot.\label{contpol} } 
\end{figure*}

The previous section has shown that the continuum polarization follows closely
the prediction of the telescope model, when no strong magnetic fields are
present. This time-dependent contribution to the continuum polarization was
thus subtracted from the observed continuum polarization. Figure \ref{contpol}
shows the resulting 2-D maps of the continuum polarization in the
observations and in the synthetic spectra, where the spatial variation of the
polarization level with the solar surface structure can be seen. I remark
once more that the continuum polarization in the synthetic spectra can only be
due to the Zeeman effect, because no other processes were considered in the
generation of the spectra with the SIR code. For relating the polarization
signal to the magnetic field, also the continuum intensity map and the field
strength map are shown. As this is the most important figure in the search for
a possibly axion-induced continuum polarization, it deserves an extended
description, starting with the circular continuum polarization ({\em 3rd column from the left}). 

For the circular continuum polarization, the observed
and synthetic polarization signals match closely both in the locations and
also in the sign of the polarization signal. The strong positive polarization
signals ({\em white}) are restricted to the umbra of the big sunspot, strong negative
polarization signals ({\em black}) appear for the small sunspot and on some
locations below it. Outside the contour lines that mark the end of the
penumbra for the two sunspots, the polarization level is close to zero. This
again indicates that the circular continuum polarization is only due to the
Zeeman effect: only for the strong fields in the sunspots the spectral lines
split\footnote{$\Delta \lambda \propto B$} enough to influence the continuum window. 

Stokes $Q$ ({\em 1st column}) shows a different behavior. All areas, which are dark (umbra of big and small sunspot) in the continuum intensity, show increased
 positive polarization signal. The penumbra shows slightly positive
 polarization signals. This does not match the polarization signal in the
 synthetic spectra ({\em top row}). There the umbra is not prominent in polarization; the
 maximum is located near the lower part of the big sunspot close to the
 penumbra. Note that the display range of the synthetic spectra ($\pm 0.05$
 \%) is smaller by a factor of 5 than that of the observed polarization ($\pm
 0.25$ \%). Outside the contour lines, the polarization level is close to zero
 everywhere as for Stokes $V$. 

Stokes $U$ ({\em 2nd column}) differs again from Stokes $Q$. The observed continuum polarization differs even stronger from the synthetic one than for Stokes $Q$. The umbra shows again strong positive polarization signal in the observed continuum polarization, but in the penumbra mainly negative
continuum polarization is seen. At the left border between the umbra and
penumbra of the big sunspot, a ridge of strong negative polarization values is
seen ({\em black color})\footnote{It creates a bit of a 3-D effect.}. Contrary to
Stokes $Q$ and $V$, the polarization signals shows a strong spatial variation
with a discernible fine structure.  Outside the contour
lines, the signal level is low, but still shows spatial structures: several
prominent app.~circular ``blobs'' of negative polarization signal can be seen
above the two sunspots. The spatial sizes of the ``blobs'' match that of
convection cells, the granulation pattern, which is seen in the intensity
map. The brightest areas show a negative polarization signal (dark in
continuum polarization), the darkest areas a positive signal (bright). This
relation between intensity structures, which are not related to magnetic fields at all, hints at the presence of cross-talk between intensity
and Stokes $U$. In total, the circular polarization signal seems to be induced
only by the Zeeman effect and is identical in synthetic and observed spectra,
but the linear polarization signal in the observed spectra strongly differs
from that of the synthetic ones, like for the average values in the previous section.
\subsection{Relation between continuum polarization and field strength}
The previous sections have shown that the observed circular continuum
polarization is dominated by the Zeeman effect. The observed linear continuum
polarization showed strong deviations from the synthetic one. If the
linear polarization signal is produced by axion-related processes that are not
 contained in the synthetic spectra, the signal level should show a dependence
 on field strength (cf.~Eq.~(\ref{eq1})) that is known from the inversion of the observed spectra.
\begin{figure}
\centerline{\resizebox{8cm}{!}{\includegraphics{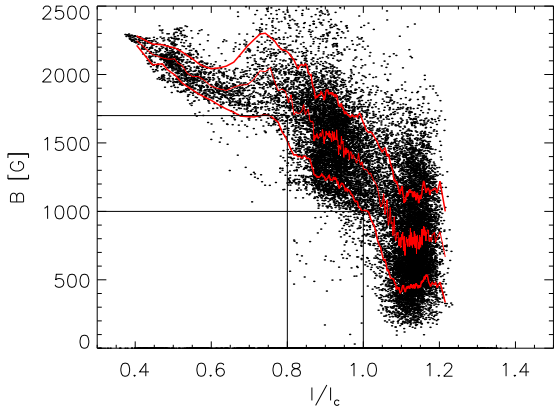}}}
\caption{Scatterplot of continuum intensity (I$_c$) vs. field strength
  (B). The {\em vertical and horizontal lines} mark two steps in the relationship: if the intensity is below 1, the fields are generally above 1000 G
  (penumbra), if the intensity is below 0.8, the fields are above 1700 G
  (umbra). The  {\em red lines} denote the average value and its standard deviation.\label{ivsb}}
\end{figure}

Before investigating the relation between field strength and polarization, I
turn to the relation between field strength and intensity because this is  a
crucial point for the later discussion. The energy transport near the solar
surface is effected by (overturning) convection rolls: hot material rises vertically
(granules), radiates energy away, moves laterally to the side, and sinks down
in cool small-scale downdrafts (intergranular lanes). The presence of
magnetic fields suppresses convection rolls; the ionized solar plasma cannot cross
the field lines. Due to buoyancy forces, solar magnetic fields are mainly
vertical to the surface. Thus, inside strong fields no lateral motion is
possible and the energy transport is strongly impeded. The
material cools and leads to the appearance of dark structures (umbra,
penumbra) whenever strong magnetic
fields are present\footnote{If the magnetic field is weak, it is swept away by
the lateral flows.}. 

Figure \ref{ivsb} shows a scatterplot of intensity vs.
field strength to visualize their relation.  The intensity was normalized to
the average intensity in the whole FOV including the dark sunspots. For the
locations outside of the sunspots (I$>$1), in principle any field strength between 0 and 1700 G can appear for a given intensity. For reduced intensities from 0.8 to 1 (penumbra)
and 0.4 to 0.8 (umbra), a minimum field strength is required (1000 G and 1700
G, respectively). Intensities below 1 cannot be achieved without magnetic fields.

\begin{figure}
\centerline{\resizebox{8cm}{!}{\includegraphics{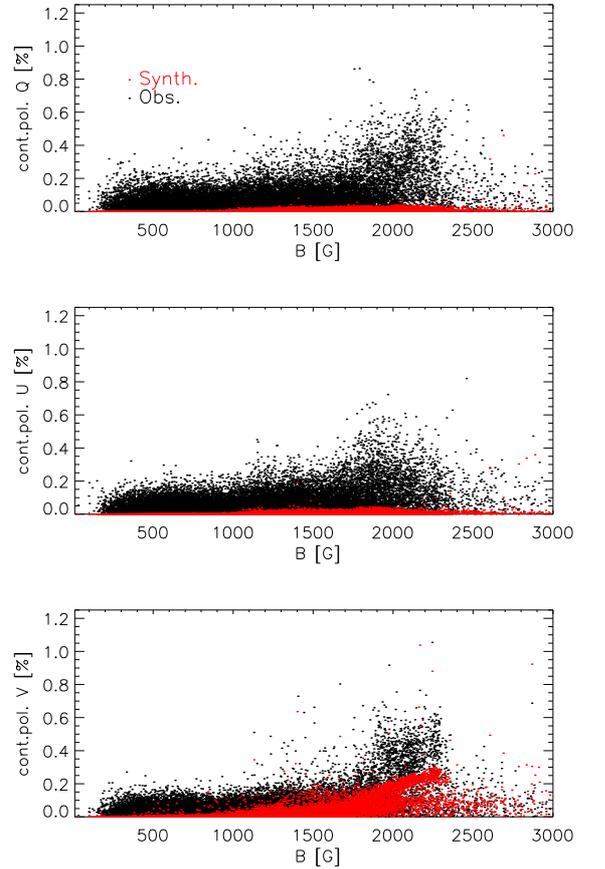}}}
\caption{Scatterplot of field strength vs. continuum polarization. {\em Top to
  bottom}: $QUV$. {\em Black}: observed spectra. {\em Red}: synthetic spectra.\label{bvspol}}
\end{figure}

Keeping the relation between intensity and field strength in mind, one can now 
turn to the scatterplot of field strength vs. continuum polarization in
Fig.~\ref{bvspol}. For each polarization ($QUV$), the observed continuum
polarization is plotted in {\em black}, together with that in the
synthetic spectra in {\em red}. Only the absolute value of the continuum
polarization is used. I remark that the synthetic spectra are free of noise in contrast to the obervations, and hence, have to show slightly lower polarization levels all the time. For Stokes $V$, again a reasonable agreement is seen:
both synthetic and observed polarization increase with field strength for field
strengths above 1500 G. For the linear polarizations $Q$ and $U$, the
synthetic polarization is always as good as zero, regardless of field
strength. The linear polarization in the observation, however, shows increased
continuum polarization signal for field strengths above 1700 G, as would be
expected from also Fig.~\ref{contpol}: the umbra has strong fields and
shows increased linear continuum polarization.
\section{Summary \& Discussion\label{summ}}
The average observed continuum polarization follows the prediction of the
telescope model closely, when no magnetic fields are present. This indicates
that the measurements can be trusted down to a polarization level of around
0.05 \%. The strong magnetic fields of the sunspots inside the field of view
show up both in the averaged continuum polarization as well as in spatially
resolved 2-D maps. The {\em circular continuum polarization} is found to be mainly
{\em due to the  Zeeman effect} in the presence of magnetic fields. In any comparison
of the observed circular continuum polarization with that in synthetic spectra
generated including only the Zeeman effect, a nearly perfect agreement is found
(Figs.~\ref{indpol} and \ref{contpol}). The change of the sign of the observed
circular continuum polarization between the big and small sunspot inside the
field of view can be directly traced back to the change of the polarity in the
Stokes $V$ spectra. 
\begin{figure}
\centerline{\includegraphics{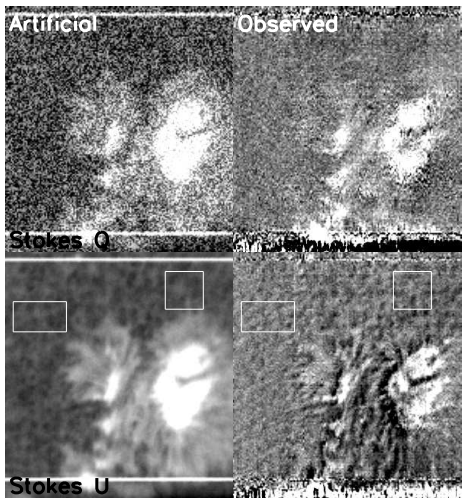}}
\caption{Comparison of artificially constructed maps of continuum polarization
  ({\em left column}) with the observed continuum polarization ({\em right column}). {\em Top row}: Stokes $Q$. {\em Bottom row}: Stokes $U$. The {\em white rectangles} outline areas, where spatial structures can be well seen in artificial and observed map.\label{artmaps}}
\end{figure}

For the linear polarization, the continuum polarization in the observed and synthetic spectra does not match. The umbra of the spots shows stronger continuum polarization signal in the observed spectra, whereas the synthetic spectra predict a maximum polarization outside of the umbra. In the observed continuum polarization of Stokes $U$, spatial structures can be discerned outside the sunspot. They resemble the granulation pattern seen in the intensity map, i.e., they are not
related to the spatial distribution of magnetic fields but of intensity. The
scatterplot of field strength vs. the continuum polarization shows an increase
of the continuum polarization for strong magnetic fields above 1700 G.

It can, however, be demonstrated that the enhanced linear polarization signal is {\em not} related to axions but residual cross-talk from intensity into the linear polarization. The appearance of the granulation/intensity pattern in the observed Stokes $U$ continuum polarization already is a clear indication of that. The prominent appearance of the sunspot(s) in the polarization maps also is no problem: the intensities in the umbra and penumbra are around 0.4 and 0.8  of the average intensity, respectively. Any linear relation between intensity and polarization thus will lead to a prominent signature of the sunspots in polarization. To demonstrate that the observed continuum polarization is due to residual cross-talk, I generated two artificial continuum polarization maps from the continuum intensity map, $I$, by:
\begin{equation}
P_{\rm art} =  -0.007 \cdot I + \epsilon \;,
\end{equation}
where $\epsilon$ is an added random noise contribution, and $P_{\rm art}$
stands for Stokes $Q$ and Stokes $U$, respectively. For Stokes $U$, the
noise was chosen to have around 4 \% of the signal amplitude, for Stokes $Q$
around 50 \%. 

\begin{figure}
\centerline{\resizebox{6cm}{!}{\includegraphics{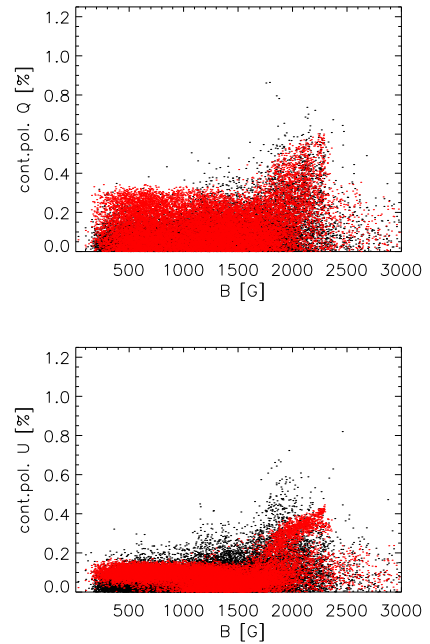}}}
\caption{Scatterplot of field strength vs. continuum polarization. {\em Top to
  bottom}: $QU$. {\em Black}: observed spectra. {\em Red}: artificial continuum polarization maps.\label{deathtoaxions}}
\end{figure}

Figure \ref{artmaps} shows the artificial continuum polarization maps thus
obtained, together with the observed continuum polarization maps. The
artificial maps show the inverted intensity structure due to the minus sign in
their generation, which was prompted by the anti-correlation between
intensity and linear continuum polarization signal The spatial structures in
the observed Stokes $U$ continuum polarization can be found one-to-one again
in the artificial map. The reduced polarization signal ({\em black}) at the
umbra-penumbra boundary at the left side of the big spot appears also in the
artificial map, even if it is less prominent there. Several of the ``blobs'' (=granules) outside the spot can be
directly identified with their counterpart in the artificial map. Two
rectangles highlight areas, where the correspondence can best be seen. In the
left one, an app.~straight line of increased polarization signal crosses the
rectangle from the lower left to the upper right corner in the artificial
map. The same line is faintly visible in the observed continuum polarization
map. In the second rectangle, a larger blob of increased signal is located
around the center. It appears again in the observed continuum polarization map.
As final test, the same scatterplot of field strength vs. continuum
polarization as in Fig.~\ref{bvspol} was done for the artificial maps
(Fig.~\ref{deathtoaxions}). Especially for Stokes $Q$, the observed and
artificial polarization signal are indistinguishable, including the increase
of polarization for large field strengths. The noise level in the artificial Stokes $Q$ map was chosen a bit too high, but that has no consequences on the conclusions.

From the comparison with the synthetic spectra and the reproduction of the spatial distribution of the linear continuum polarization by cross-talk from $I$ into $Q$ and $U$, I would exclude the presence of excess unexplained (possibly axion-related) polarization signal above a level of around  0.05 \% of $I_c$, or 5 out of 10000 photons. This limit in polarization degree translates, however, only into a very weak upper limit for the axion-photon coupling due to the large number of photons that are present from the solar black-body radiation. The number of axion-related photons, $\gamma_{ax}$, would be given by
\begin{equation}
\gamma_{ax} = p_{a\rightarrow\gamma} n_{ax} \;,
\end{equation}
where $n_{ax}$ is the number of axions at the solar surface. \citet{andriamonje+etal2007} give a value of around $\phi_a = g_{10}\cdot 10^{15}$ m$^{-2}$s$^{-1}$keV$^{-1}$ for the solar surface axion flux, with $g_{10} = g_{a\gamma}/ 10^{-10}$ GeV$^{-1}$.

Using the approximation of \citet{cameron+etal1993}, $p_{a\rightarrow\gamma}$ is proportional to $(g_{a\gamma}BL)^2$. The continuum radiation comes only from a small layer in the solar atmosphere with around 50-200 km vertical
extension. Using an upper limit of $B=0.2$ T and $L=$ 200 km, $(BL)^2 \sim
10^{9}$ (Tm)$^2$. This gives an axion-related photon flux, $\phi_{\gamma,a}$ of 
\begin{equation}
\phi_{\gamma,a} = \phi_a \cdot p_{a\rightarrow\gamma} \sim (g_{10})^3 \cdot 10^{24} {\rm m}^{-2}{\rm s}^{-1}{\rm keV}^{-1} \;.
\end{equation}
The additional polarization signal would be given by the fraction of axion-related photon flux to that from the unpolarized black-body ($bb$) radiation, $\gamma_{bb}$. With a solar surface temperature of 5800 K, the Planck curve predicts a power of $p=7.5\cdot 10^{10}$ Wm$^{-2}$nm$^{-1}$ near 630 nm. The continuum window used extends over around 0.01 nm, and the energy of a single photon at 630 nm is around 3$\cdot 10^{-19}$ J ($\sim2$eV). The radiation power then transforms into a photon flux from the black-body radiation of 
\begin{equation}
\phi_{\gamma,bb} = 7.5\cdot 10^{10}\cdot 0.01 / 3 \cdot 10^{-19} = 2.4\cdot 10^{27} {\rm m}^{-2}{\rm s}^{-1} \;.
\end{equation}
Assuming that the axion-flux in the eV range is at least an order of magnitude lower than near its maximum at around 3 keV \citep[][their Fig.~2]{andriamonje+etal2007} and transforming the extension of the continuum window in the spectrum into an energy interval of 3$\cdot 10^{-5}$ eV, yields 
\begin{equation}
\phi_{\gamma,a}= (g_{10})^3 \cdot 3\cdot 10^{15} {\rm m}^{-2}{\rm s}^{-1} \;.
\end{equation}
With the observed upper limit that less than 5$\cdot 10^{-4}$ of the photons in the continuum window show additional polarization signal from possibly axion-related processes, it follows that 
\begin{eqnarray}
(g_{10})^3 &&< 5\cdot 10^{-4} \cdot  2.4\cdot 10^{27} / 3\cdot 10^{15} = 4\cdot 10^8 \nonumber\\
\mbox{or } g_{10} &&< 700 \; .
\end{eqnarray}
\section{Conclusion}
The final result of this investigation is a negative one: there are no traces
for axion-induced polarization signals in a continuum wavelength near 630
nm above a level of around 0.05 \% of $I_c$.  The residual linear continuum polarization signal at locations of strong magnetic fields could be shown to be due to cross-talk from intensity into polarization. This leads\footnote{In this estimate it is assumed that an axion of energy $E$ produces a photon of the same energy, but I don't know if that is to be expected. I'm also not fully sure about units, but \citet{cameron+etal1993} give  $(g_{a\gamma} BL)^2$ as a unitless probability. I hope this comes out naturally using SI units for $B$ and $L$.} to a value of the axion-photon coupling $g_{a\gamma} < 700 \cdot 10^{-10}$ GeV$^{-1}$. \citet{andriamonje+etal2007} exclude a value of $g_{a\gamma} > 4.5\cdot 10^{-10}$ GeV$^{-1}$, which is some orders of magnitude below the present estimate. The present method of estimating $g_{a\gamma}$ could be improved by observations with a significant better S/N ratio, produced by much longer exposure times, or large-scale spatial averaging over, e.g., the whole umbra and penumbra of a spot. In addition, the active region was off the center of the solar disk by $>$10\,degrees. Solar axions should only induce real photons in forward scattering, so observations of an active region closer to or best directly at disc center would be preferred. 
\begin{acknowledgements}
I thank K.Zioutas for providing the idea for this study. The VTT is operated by the Kiepenheuer-Institut f\"ur Sonnenphysik (KIS) at the Spanish Observatorio del Teide of the Instituto de Astrof\'{\i}sica de Canarias (IAC). The POLIS instrument has been a joint development of the High Altitude Observatory (Boulder, USA) and the KIS.
\end{acknowledgements}
\bibliographystyle{aa}
\bibliography{references}

\end{document}